\pdfoutput=1

\documentclass[showpacs,aps,prl,twocolumn,superscriptaddress]{revtex4}
\usepackage{graphicx} 
\usepackage{dcolumn}
\usepackage{bm}
\usepackage{amssymb,amsmath}
\usepackage{epstopdf}
\usepackage[T1]{fontenc}
\usepackage[latin9]{inputenc}
\usepackage{color}
\usepackage{float}
\usepackage{esint}
\usepackage{graphics}
\makeatletter
\@ifundefined{textcolor}{}
{%
 \definecolor{BLACK}{gray}{0}
 \definecolor{WHITE}{gray}{1}
 \definecolor{RED}{rgb}{1,0,0}
 \definecolor{GREEN}{rgb}{0,1,0}
 \definecolor{BLUE}{rgb}{0,0,1}
 \definecolor{CYAN}{cmyk}{1,0,0,0}
 \definecolor{MAGENTA}{cmyk}{0,1,0,0}
 \definecolor{YELLOW}{cmyk}{0,0,1,0}
}

\begin{document}
\title{Stability of High-Density Two-Dimensional Excitons against a Mott Transition \\in High Magnetic Fields Probed by Coherent Terahertz Spectroscopy}
\normalsize

\author{Qi Zhang}
\affiliation{Department of Electrical and Computer Engineering, Rice University, Houston, Texas 77005, USA}

\author{Yongrui Wang}
\affiliation{Department of Physics and Astronomy, Texas A\&M University, College Station, Texas 77843, USA}

\author{Weilu Gao}
\affiliation{Department of Electrical and Computer Engineering, Rice University, Houston, Texas 77005, USA}

\author{Zhongqu Long}
\affiliation{Department of Physics and Astronomy, Texas A\&M University, College Station, Texas 77843, USA}

\author{John D.~Watson}
\affiliation{Department of Physics and Astronomy, Station Q Purdue, and Birck Nanotechnology Center, Purdue University, West Lafayette, Indiana 47907, USA}

\author{Michael J.~Manfra}
\affiliation{Department of Physics and Astronomy, Station Q Purdue, and Birck Nanotechnology Center, Purdue University, West Lafayette, Indiana 47907, USA}
\affiliation{School of Materials Engineering and School of Electrical and Computer Engineering, Purdue University, West Lafayette, Indiana 47907, USA}

\author{Alexey Belyanin}
\affiliation{Department of Physics and Astronomy, Texas A\&M University, College Station, Texas 77843, USA}

\author{Junichiro Kono}
\email[]{kono@rice.edu}
\affiliation{Department of Electrical and Computer Engineering, Rice University, Houston, Texas 77005, USA}
\affiliation{Department of Physics and Astronomy, Rice University, Houston, Texas 77005, USA}
\affiliation{Department of Materials Science and NanoEngineering, Rice University, Houston, Texas 77005, USA}

\date{\today}

\begin{abstract}
We have performed time-resolved terahertz absorption measurements on photoexcited electron-hole pairs in undoped GaAs quantum wells in magnetic fields.  We probed both unbound- and bound-carrier responses via cyclotron resonance and intraexciton resonance, respectively. The stability of excitons, monitored as the pair density was systematically increased, was found to dramatically increase with increasing magnetic field. Specifically, the 1$s$--2$p_-$ intraexciton transition at 9\,T persisted up to the highest density, whereas the 1$s$--2$p$ feature at 0\,T was quickly replaced by a free-carrier Drude response.  Interestingly, at 9\,T, the 1$s$--2$p_-$ peak was replaced by free-hole cyclotron resonance at high temperatures, indicating that 2D magnetoexcitons do dissociate under thermal excitation, even though they are stable against a density-driven Mott transition. 
\end{abstract}


\maketitle

Electron-hole ($e$-$h$) pairs in solids provide highly controllable two-component systems with a variety of possible phases brought about by long-range Coulomb interactions~\cite{Jeffries75Science,GriffinetAl95Book}.  Density-dependent Coulomb interactions can drive $e$-$h$ pairs through an excitonic Mott transition from an insulating excitonic gas into a metallic $e$-$h$ plasma. Theoretical studies have suggested that these interactions can be substantially modified by a magnetic field ($B$), creating {\em magnetoexcitons}. At low $e$-$h$ pair densities, the principal change introduced by $B$ is the mixing of the center-of-mass (COM) and relative coordinates~\cite{Knox63Book,GorkovDzyaloshinskii67JETP}, which leads to nonmonotonic energy dispersions~\cite{LernerLozovik1980JETP}. In addition, the Mott density is naturally expected to increase with $B$ due to the shrinkage of exciton wavefunctions. The most striking theoretical prediction is that exciton-exciton interactions completely vanish in 2D systems in a strong perpendicular $B$ due to an $e$-$h$ charge symmetry, the so-called hidden symmetry, which results in an exact cancellation of Coulomb interactions between excitons, making 2D magnetoexcitons ultrastable~\cite{MacDonaldRezayi90PRB,DzyubenkoLozovik91JPA,ApalkonRashba91JETP}. A previous experimental study showed that a hidden symmetry phase appears abruptly at a filling factor of exactly two~\cite{YoonetAl97SSC}, namely, when all carriers are accommodated by the lowest Landau level.

High-density magnetoexcitons in semiconductor quantum wells (QWs) were previously studied through continuous-wave (CW) spectroscopy~\cite{PotemskietAl90SSC,ButovetAl92PRB}, while time-resolved studies are limited~\cite{StarketAl90PRL,KochetAl99SSC,ButovetAl01PRL,LozoviketAl02PRB}, although they are more desired for studying exciton dynamics and Mott physics, since much higher exciton densities can be achieved.  Moreover, most previous studies employed interband photoluminescence and absorption spectroscopy for probing excitons. However, there are two inherent limitations in utilizing interband optical transitions for monitoring exciton dynamics. First, due to the negligibly small momentum of photons, excitons with nonzero COM momenta, $P_\mathrm{COM}$, are dark (or inaccessible) for interband transitions. Second, it is difficult to clearly identify spectral features for bound and unbound carriers; nonexcitonic effects can also affect interband processes, including bandgap renormalization (BGR). These limitations are not pertinent to {\em intraexciton} spectroscopy using terahertz (THz) radiation~\cite{CerneetAl96PRL,SalibetAl96PRL,KonoetAl97PRL}, which conserves the number of excitons while allowing one to access dark exciton states with finite $P_\mathrm{COM}$. Also, intraexciton transitions are essentially uninfluenced by BGR. For bound and unbound carriers, THz spectroscopy reveals distinctly different features, i.e., intraexciton and Drude responses, whose characteristics can provide quantitative information on their properties.  Hence, THz time-domain spectroscopy has been used for studying excitonic dynamics and Mott physics in various semiconductors~\cite{KaindletAl03Nature,HuberetAl05PRB,Lloyd-HughesetAl08PRB,SuzukiShimano09PRL,SuzukiShimano12PRL,RiceetAl13PRL,BhattacharyyaetAl14PRB,SekiguchiShimano15PRB,LuoetAl15PRL,PollmannetAl15NM}. 

Here, we present results of a THz study of $e$-$h$ pairs in photoexcited undoped GaAs QWs in $B$, using optical-pump/THz-probe spectroscopy. We simultaneously monitored the intraexcitonic 1$s$--2$p$ transition (which splits into the 1$s$--2$p_+$ and 1$s$--2$p_-$ transitions in $B$) and the cyclotron resonance (CR) of unbound electrons and holes as a function of $B$, temperature ($T$), $e$-$h$ pair density, and time delay ($t$).  The 1$s$--2$p_-$ feature was robust at high $B$ even at high densities, whereas the 1$s$--2$p$ feature at 0\,T quickly disappeared, and a free-carrier Drude response emerged at high densities.  Interestingly, the 1$s$--2$p_-$ peak at 9\,T was replaced by hole CR at high $T$, suggesting that thermal dissociation of 2D magnetoexcitons can occur despite the fact that they are extremely stable against a density-driven Mott transition into an $e$-$h$ magnetoplasma. Our theoretical model incorporating the $e$-$h$ Coulomb interaction non-perturbatively successfully explains the $T$-dependence of THz spectra at 9\,T, but not the density dependence.  We discuss these results in light of the hidden symmetry of 2D magnetoexcitons~\cite{DzyubenkoLozovik91JPA,MacDonaldRezayi90PRB,ApalkonRashba91JETP}.

\begin{figure}
\includegraphics[scale=0.46]{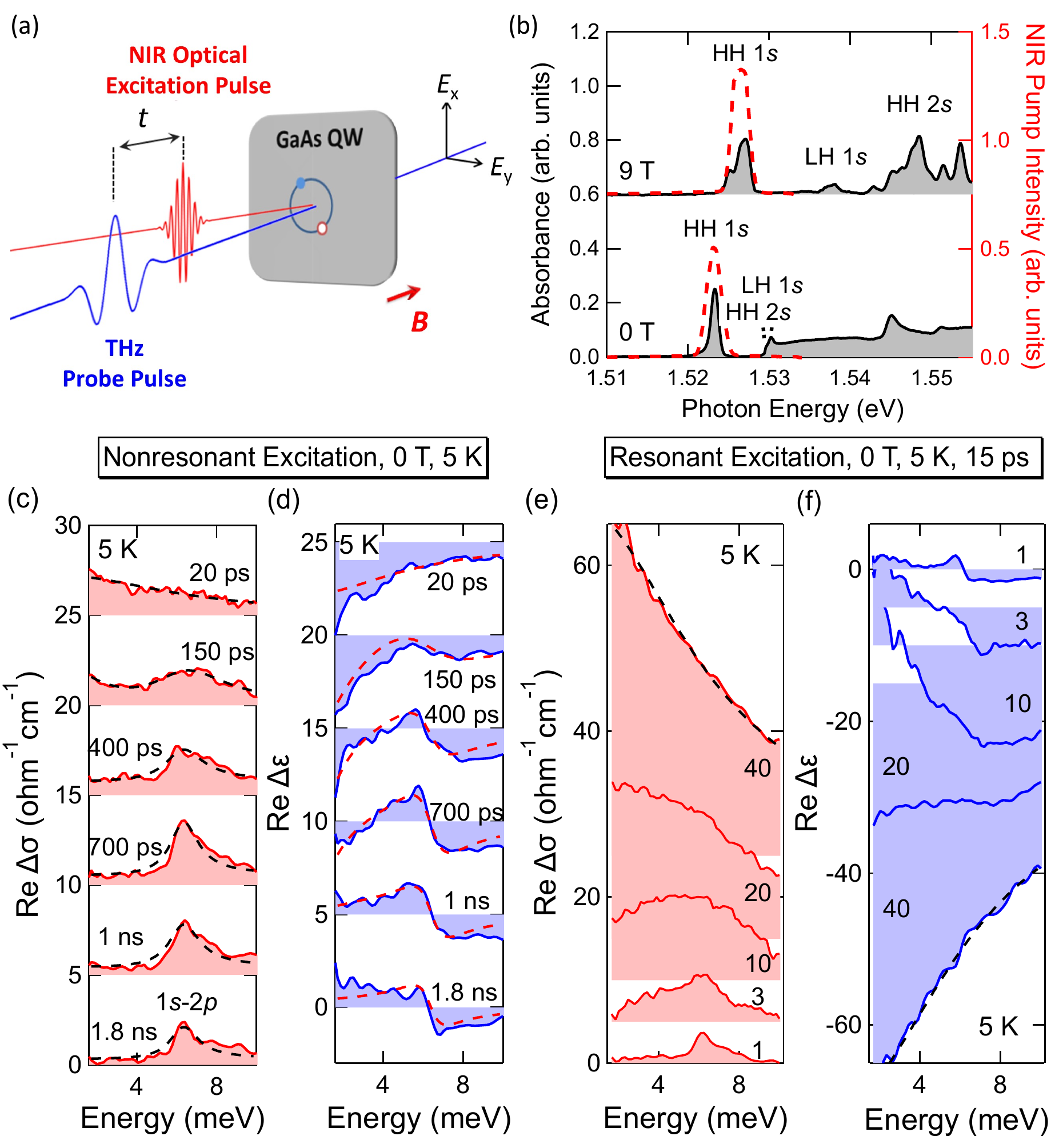}
\caption{(a)~Schematic diagram of the experimental setup for measuring exciton dynamics in undoped GaAs quantum wells in a magnetic field using optical-pump/THz-probe spectroscopy. (b)~Resonant pump spectrum (dashed line), together with a linear absorption spectrum (solid line) for the sample, at 0\,T (bottom) and 9\,T (top). (c),~(d)~Evolution of THz spectra after nonresonant excitation at 0\,T with a fluence of 150\,nJ/cm$^{2}$. Photoninduced change in the real part of the (c)~conductivity, $\Delta\sigma(\omega)$, and (d)~permittivity, $\Delta\varepsilon(\omega)$, at 0\,T.  At 20\,ps, $\mathrm{Re}\{\Delta\varepsilon(\omega)\}$ is negative, a characteristic metallic behavior. The 1$s$--2$p$ transition is observed at 6.41\,meV after 150\,ps and survives even at 1.8\,ns.  (e),~(f)~$\mathrm{Re}\{\Delta\sigma(\omega)\}$ and $\Delta\mathrm{Re}\{\varepsilon(\omega)\}$ with various resonant pump fluences at 0\,T and 15\,ps. The pump fluence from the bottom to top ranges from 1\,$\times$\,150\,nJ/cm$^{2}$ to 40\,$\times$\,150\,nJ/cm$^{2}$. An excitonic Mott transition is observed: The 1$s$--2$p$ transition vanishes at the highest pump fluence, where a Drude-like spectrum appears. Dashed lines in (c)-(f) are fits using Eqs.\,(4)-(6) in Supplemental Material~\cite{SM}.}
\label{setup-etc}
\end{figure}

Experiments were carried out on an undoped GaAs QW sample, containing 15 pairs of a 20-nm GaAs QW and a 20-nm Al$_{0.3}$Ga$_{0.7}$As barrier, sandwiched by two 500-nm-thick Al$_{0.3}$Ga$_{0.7}$As spacer layers. The total thickness of the multiple QW layers was 600\,nm. The GaAs substrate was removed by selective etching to avoid any photoinduced carrier effects in bulk GaAs. We used a Ti:Sapphire regenerative amplifier (1\,kHz, 150\,fs, 775\,nm, Clark-MXR, Inc.)~ to generate and detect THz pulses with ZnTe crystals. As schematically shown in Fig.\,\ref{setup-etc}(a), an optical excitation pulse created heavy-hole (HH) $1s$ excitons (unbound $e$-$h$ pairs) in the QWs through resonant (nonresonant) excitation, and a THz probe pulse transmitted through the sample at a certain time delay, $t$. For resonant excitation, a $4f$ pulse shaper was used as a tunable ultranarrow bandpass filter to select the frequency component that was resonant with the HH $1s$ state from an optical parametric amplifier, as shown in Fig.\,\ref{setup-etc}(b). In the case of nonresonant excitation, the sample was excited by a 1.6\,eV pulse, $\sim$70\,meV higher than the HH $1s$ energy, thus creating a hot $e$-$h$ plasma in the continuum.  At fixed values of $t$, we recorded the optical-pump-induced change of the transmitted THz field, $\Delta{E}(t)$. Combining it with a separate reference measurement without optical excitation, $E_\mathrm{ref}(t)$, we obtained the photoinduced change in the (longitudinal) complex permittivity, $\Delta\varepsilon(\omega)$, and (longitudinal) complex optical conductivity, ${\Delta}\sigma(\omega)$~\cite{SM}.


In the case of nonresonant excitation [Figs.\,\ref{setup-etc}(c) and \ref{setup-etc}(d)], a hot distribution of $e$-$h$ pairs is created at $t =$ 0 by a 1.6\,eV pump pulse with a fluence of 150\,nJ/cm$^{2}$ at 5\,K and 0\,T. At $t =$ 20\,ps, $\Delta\varepsilon(\omega)$ is negative, and no exciton feature is observed, and thus, the system is metallic. As time progresses, the 1$s$--2$p$ transition emerges at 6.41\,meV at 150\,ps and becomes clearer at 400\,ps; it eventually dominates the conductivity spectrum at later times, signaling the completion of cooling and exciton formation dynamics~\cite{KaindletAl03Nature}. Exciton population then gradually decreases through interband recombination, but some excitons survive even at $t =$ 1.8\,ns. The measured THz 1$s$--2$p$ transition energy matches well the 6.4\,meV energy separation between the HH 1$s$ and 2$s$ states measured by CW absorption spectroscopy [see Fig.\,\ref{setup-etc}(b)]. The corresponding exciton binding energy, $E_\mathrm{b}$, is 7.2\,meV within the 2D hydrogen model.

%
%

Figures~\ref{setup-etc}(e) and \ref{setup-etc}(f) present the pump fluence dependence of $\Delta\sigma(\omega)$ for resonant excitation at $B =$ 0\,T, $T =$ 5\,K, and $t =$ 15\,ps. At low fluences, $\Delta\sigma(\omega)$ shows a pronounced peak due to the 1$s$--2$p$ transition.  As the density of HH 1$s$ excitons increases with increasing pump fluence, this feature broadens very quickly, while, at the same time, a Drude-like response emerges in the low frequency region. Eventually, at the highest pump fluence (40\,$\times$\,150\,nJ/cm$^{2}$), the 1$s$--2$p$ transition totally vanishes, and the system is a correlated $e$-$h$ plasma. The carrier scattering time and $e$-$h$ pair density were extracted by Drude fitting of the highest fluence trace. The obtained scattering time was 0.1\,ps, and the density was 2.86\,$\times$\,10$^{12}$\,cm$^{-2}$, which corresponds to 1.9\,$\times$\,10$^{11}\,$cm$^{-2}$ per QW. Therefore, in our system, the Mott density is on the order of 10$^{11}$\,cm$^{-2}$ for a single 20-nm-wide QW, which is similar to a reported value~\cite{HuberetAl05PRB}. 

\begin{figure}
\includegraphics[scale=0.41]{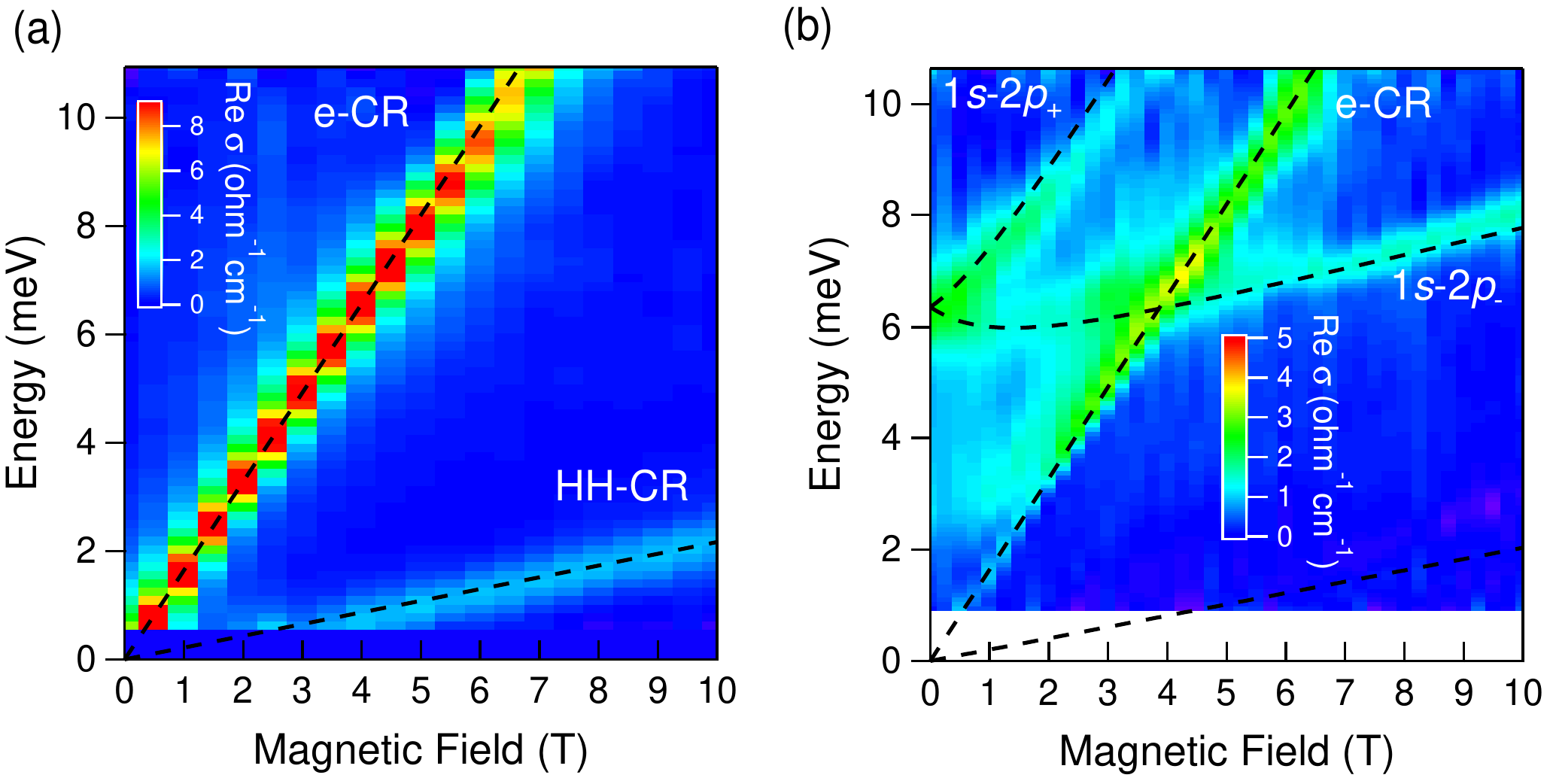}
\caption{Photoinduced conductivity, $\mathrm{Re}\{\Delta\sigma(\omega)\}$, as a function of magnetic field from 0\,T to 10\,T at a time delay of (a)~15\,ps and (b)~1\,ns after nonresonant excitation. At 15\,ps, a hot $e$-$h$ magnetoplasma shows electron and heavy-hole cyclotron resonance.  At 1\,ns, the intraexciton 1$s$--2$p_+$ and 1$s$--2$p_-$ transitions are observed. Dashed lines are calculated transition energies with experimental parameters.}
\label{map}
\end{figure}

A magnetic field lifts the degeneracy of the 2$p$ states, which have nonzero orbital angular momenta. Hence, the 1$s$--2$p$ transition splits into 1$s$--2$p_+$ and 1$s$--2$p_-$ transitions in $B$. In the Landau level (LL) picture, the 1$s$--2$p_+$ (1$s$--2$p_-$) transition can be viewed as a transition where the electron (hole) is excited from the lowest to the first-excited LL while the hole (electron) stays in the lowest LL. We measured $\Delta \sigma(\omega)$ at various $B$ from 0\,T to 10\,T at 5\,K. As shown in Fig.\,\ref{map}(a), after nonresonant excitation, two distinct CR features due to unbound electrons and HHs are observed for an $e$-$h$ magnetoplasma at $t =$ 15\,ps. The extracted effective masses of electrons and HHs are 0.070$m_\mathrm{e}$ and 0.54$m_\mathrm{e}$, respectively, where $m_\mathrm{e} =$ 9.11\,$\times$\,10$^{-31}$\,kg. At $t =$ 1\,ns [Fig.\,\ref{map}(b)], magnetoexcitons are already formed, and thus, we observe the 1$s$--2$p_+$ and 1$s$--2$p_-$ transitions, in addition to electron CR. 

\begin{figure} 
\includegraphics[scale=0.49]{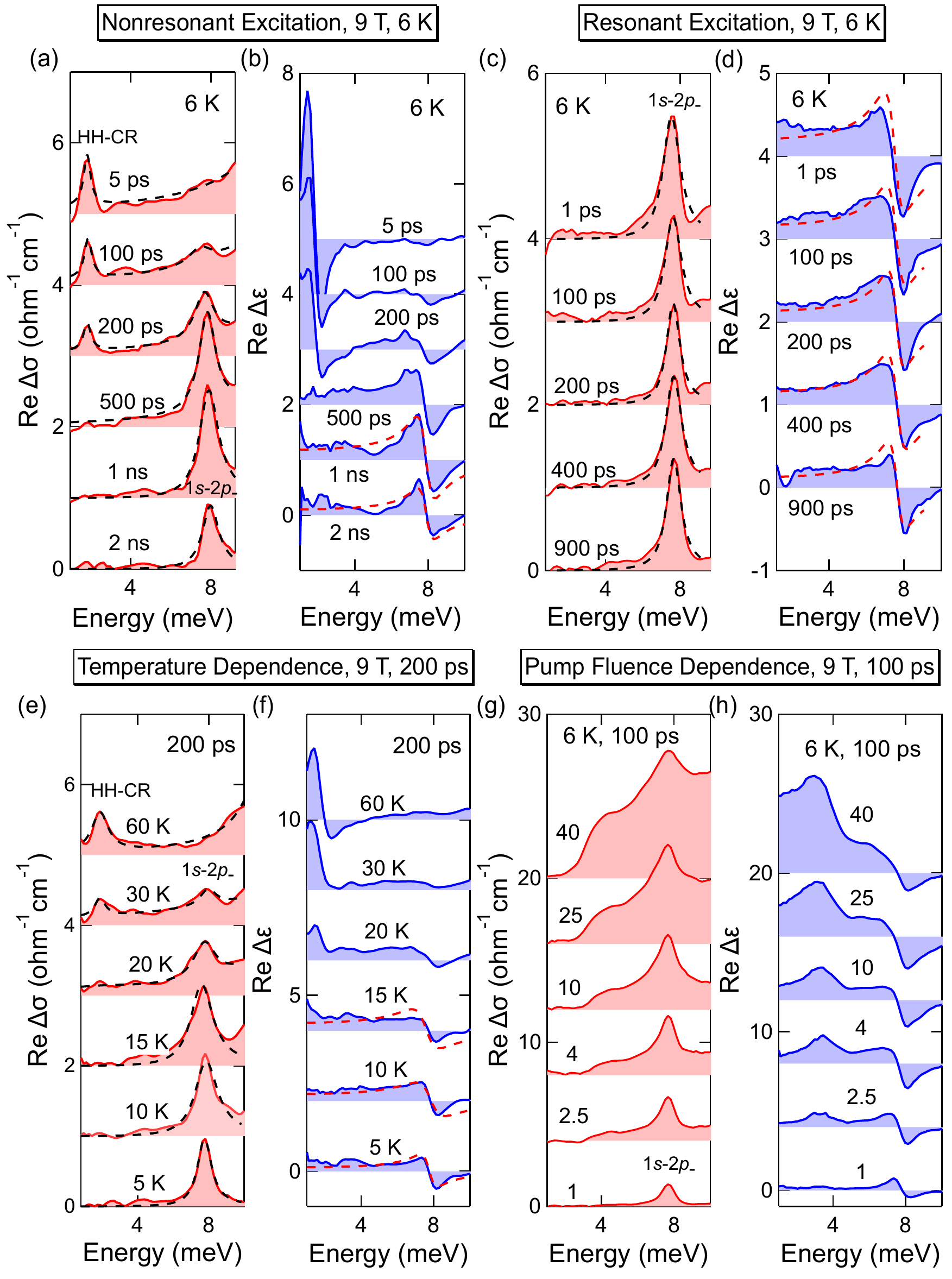}
\caption{(a)~Conductivity and (b)~permittivity spectral evolution after {\em nonresonant} excitation with a fluence of 200\,nJ/cm$^{2}$ at 9\,T. The spectra are dominated by HH-CR at early times and by the 1$s$--2$p_-$ transition at later times.  (c)~Conductivity and (d)~permittivity spectral evolution after {\em resonant} excitation of heavy-hole 1$s$ excitons with a fluence of 200\,nJ/cm$^{2}$ at 9\,T.  The 1$s$--2$p_-$ transition is the only feature observed at all time delays.  (e),~(f)~Temperature dependent conductivity and permittivity spectra at 9\,T and 200\,ps after resonant ecitation.  The 1$s$--2$p_-$ transition vanishes at high temperatures where HH-CR emerges.  (g),~(h)~Pump fluence dependent conductivity and permittivity spectra at 9\,T, 5\,K, and 100\,ps after resonant excitation.  The pump fluence varies from 1\,$\times$\,200\,nJ/cm$^{2}$ to 40\,$\times$\,200\,nJ/cm$^{2}$. The 1$s$--2$p_-$ peak survives even at the highest fluence, which would be sufficiently high to drive an excitonic Mott transition at 0\,T.}
\label{various}
\end{figure}

Figures~\ref{various}(a)-\ref{various}(d) present the time evolution of $\Delta \sigma(\omega)$ and $\Delta\varepsilon(\omega)$ at 9\,T after excitation with a fluence of 200\,nJ/cm$^{2}$. For the case of nonresonant excitation [(a) and (b)], magnetoexciton formation is observed.  Namely, HH-CR ($\sim$2\,meV) and the low-energy tail of electron CR (which occurs outside our bandwidth) are observed at $t =$ 5\,ps. However, they decay in time and are eventually replaced by the 1$s$--2$p_-$ transition after 100\,ps. This behavior should be contrasted to the case of resonant excitation [(c) and (d)], in which the 1$s$--2$p_-$ transition immediately appears after excitation and continues to be the only observable feature throughout the time delay range up to 900\,ps while its intensity gradually decreases with time through interband recombination.

Thermal ionization of resonantly created magnetoexcitons is presented in Figs.\,\ref{various}(e) and \ref{various}(f). HH-CR and electron CR increase in intensity with increasing $T$ at the expense of the intensity of the 1$s$--2$p_-$ transition. At 60\,K, the exciton feature completely vanishes, which is consistent with the fact that the thermal energy at 60\,K, 5.25\,meV (or 1.27\,THz), is close to the exciton binding energy. Figures~\ref{various}(g) and \ref{various}(h) illustrate the pump fluence dependence of $\Delta\sigma(\omega)$ and $\Delta\varepsilon(\omega)$ at 9\,T, which is strikingly different from the $T$-dependence in (e) and (f). The 1$s$--2$p_-$ peak increases both in intensity and width with increasing fluence but remains observable even at the highest fluence (40\,$\times$\,200\,nJ/cm$^{2}$) with no apparent peak shift; {\em no CR is observed at any fluence}.  The highest fluence, corresponding to a pair density of 10$^{11}$\,cm$^{-2}$ per QW, would have driven an excitonic Mott transition at 0\,T [Figs.\,\ref{setup-etc}(e) and \ref{setup-etc}(f)].   These observations highlight an unusual aspect of 2D magnetoexcitons: while they are magnetically stabilized against a density-driven Mott transition into an $e$-$h$ magnetoplasma, they are still unstable against thermal ionization.

\begin{figure} 
\centering
\includegraphics[width=0.95\linewidth]{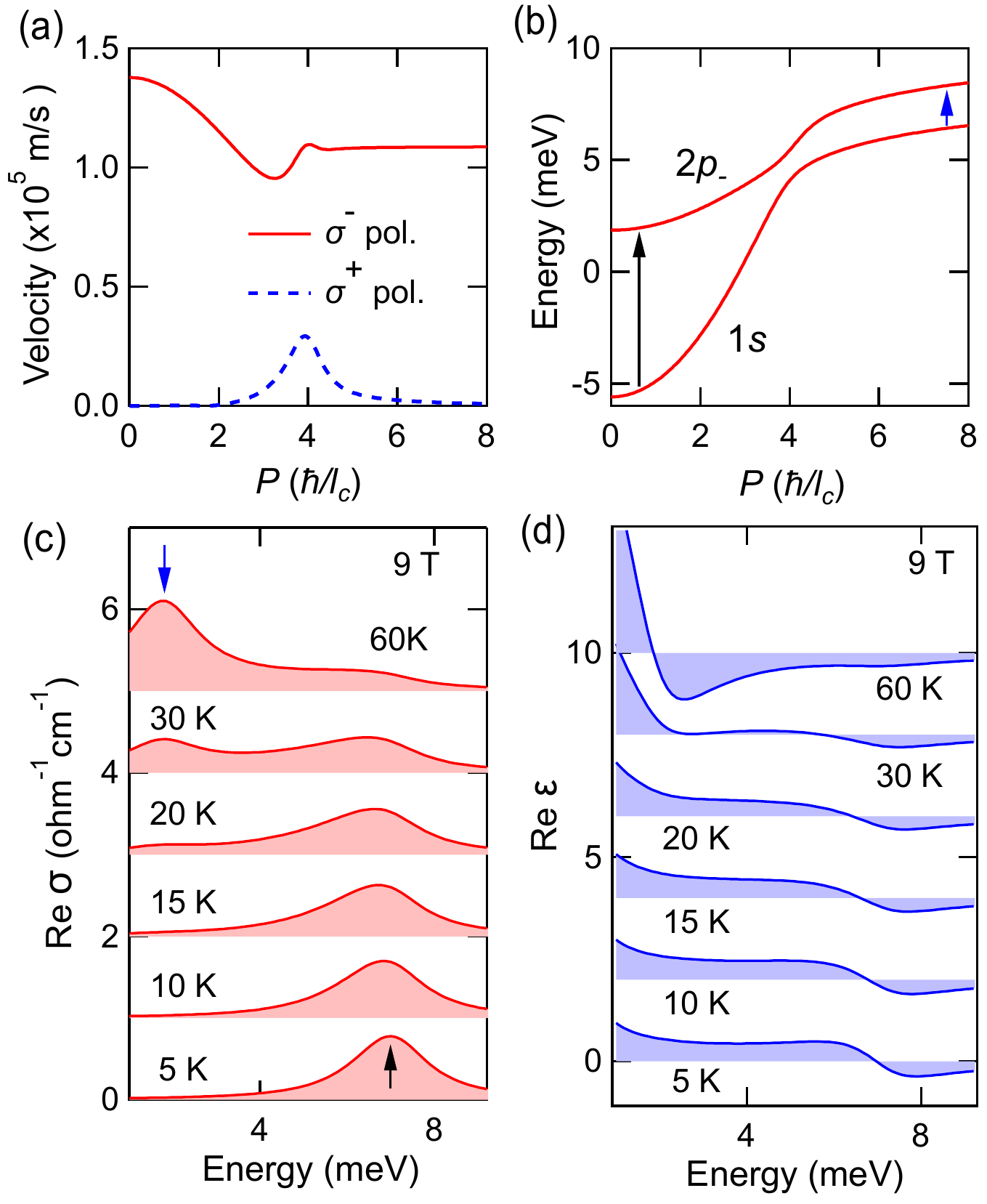}
\caption{(a) The dependence of the velocity matrix elements on the COM momentum, $P$, for the lowest two excitonic subbands (1$s$ and 2$p_-$) for $\sigma^+$ and $\sigma^-$ circular polarizations at 9\,T.  $l_c = \sqrt{\hbar c / e B}$ is the magnetic length. (b)~The energy dispersions for the 1$s$ and 2$p_-$ exciton subbands at 9\,T. (c) and (d): Calculated real part of THz conductivity (c) and permittivity (d) spectra for quantum-well magnetoexcitons at different temperatures for linear polarization at 9\,T.  As the temperature increases, the 1$s$--2$p_-$ transition at large $P$ gradually becomes the dominant feature of the spectrum, which corresponds to HH-CR. These calculated spectra qualitatively reproduce the experimental spectra in Figs.~\ref{various}(e) and \ref{various}(f). 
}
\label{theory}
\end{figure}

The filling factor corresponding to the highest density (10$^{11}$\,cm$^{-2}$ per QW) at 9\,T is still less than 2, so the system is in the magnetic quantum limit, where the hidden symmetry of 2D magnetoexcitons~\cite{DzyubenkoLozovik91JPA,MacDonaldRezayi90PRB,ApalkonRashba91JETP} is expected to prevent density-driven dissociation. Therefore, our observations provide evidence for the existence of the hidden symmetry in this system. To provide quantitative interpretation of the observed excitonic response,  we developed a nonperturbative two-body model for magnetoexcitons in a QW and calculated THz spectra at different $B$, $T$, and pair densities~\cite{SM}. Advancing beyond the previous perturbative approach~\cite{LozovikRuvinskii97JETP}, here we treat both the Coulomb and magnetic terms nonperturbatively for arbitrary center-of-mass momenta. This allowed us to cover the whole range from zero to strong magnetic fields and to interpret the  thermal ionization of magnetoexcitons observed in their absorption spectra using the magnetoexciton dispersion curves in Fig.\,\ref{theory}. On the other hand, our theory does not include many-body interexciton interactions and therefore would not be able to describe the density-driven Mott transition at even higher densities (i.e., at filling factors larger than 2).

The Hamiltonian is
\begin{align}
\hat{H} = & \frac{1}{2 m_e} \left( -i\hbar \bm{\nabla}_e + \frac{e}{c} \bm{A}_e \right)^2
+ \frac{1}{2 m_h} \left( -i\hbar \bm{\nabla}_e - \frac{e}{c} \bm{A}_h \right)^2  \nonumber \\
+ & V_\mathrm{C}(\bm{r}_e -\bm{r}_h),
\end{align}
where $\bm{r}_{e}$ and $\bm{r}_{h}$ are the in-plane positions of the electron and the hole, respectively, $m_{e,h}$ are their effective masses, $\bm{A}_{e,h}$ are the vector potentials, which can be chosen to be $\bm{A}_{e,h} = (1/2) \bm{B} \times \bm{r}_{e,h}$, and $V_\mathrm{C} (\bm{r}_e -\bm{r}_h)$ is the effective Coulomb potential given in~\cite{SM}. 
%
%
%
The wavefunction is
\begin{equation}
\Psi(\bm{r}_e,\bm{r}_h) = e^{\left[ \frac{i}{\hbar} \bm{R} \cdot \left( \bm{P} + \frac{e}{2c} \bm{B} \times \bm{r} \right) \right]} e^{\left( \frac{1}{2} i \gamma \bm{r} \cdot \bm{P} \right)} \Phi(\bm{r} - \bm{\rho}_0) ,
\end{equation}
where $\bm{R} = (m_e \bm{r}_e + m_h \bm{r}_h) / (m_e+m_h)$, $\bm{r} = \bm{r}_e - \bm{r}_h$, $\gamma = (m_h-m_e)/(m_h+m_e)$, $\bm{P}$ is an analogue to $\bm{P}_\mathrm{COM}$ in the absence of $B$, and $\bm{\rho}_0 = c \bm{B}\times\bm{P}/eB^2$. The wavefunctions $\Phi(\bm{r} - \bm{\rho}_0)$ are solutions of
\begin{align}
\label{Eq:eigeneq_exciton}
&\left[ -\frac{\hbar^2}{2\mu} \Delta + \frac{e\hbar}{2i\mu c} \gamma \bm{B} \cdot \bm{r} \times \bm{\nabla}
+ \frac{e^2}{8\mu c^2} B^2 r^2 + V_C(\bm{r}+\bm{\rho}_0) \right] \nonumber \\
&\times \Phi(\bm{r}) = E \Phi(\bm{r}) ,
\end{align}
which can be solved by  using the basis wavefunctions in the absence of $V_\mathrm{C}$~\cite{SM}.

The interaction of the $e$-$h$ pair and a THz field, $\bm{E} = -(1/c) \partial \bm{A} / \partial t$, can be written as $\hat{H}_\mathrm{I} = (e/c) (\bm{v}_e-\bm{v}_h) \cdot \bm{A}$, 
where $\bm{v}_e = \left(-i\hbar\bm{\nabla}_e + (e/2c) \bm{B}\times\bm{r}_e \right)/m_e$ and $
\bm{v}_h = \left(-i\hbar\bm{\nabla}_h - (e/2c) \bm{B}\times\bm{r}_h \right)/m_h$
are velocity operators. 
%
%
%
%

Assuming that the lowest exciton subband (``$1s$'') $|\Psi_1\rangle$ is occupied, the absorption spectrum for transitions between this and the second-lowest subband (``$2p_-$'') $|\Psi_2\rangle$ can be calculated according to the Kubo formula
\begin{align}
\sigma_{ij}(\omega) = &\frac{2 i e^2}{(2\pi\hbar)^2} \int d \bm{P} f_1(\bm{P}) \nonumber \\
&\times \frac{  \langle \Psi_1 | (\bm{v}_e-\bm{v}_h)_i | \Psi_2 \rangle \langle \Psi_2 | (\bm{v}_e-\bm{v}_h)_j | \Psi_1 \rangle  }{\left[\hbar\omega - (E_2-E_1) + i\hbar\delta \right](E_2-E_1)}.
\end{align}
Here, $\delta$ is the broadening factor, $f_1(\bm{P})$ is the Bose-Einstein distribution function for excitons in the $1s$ excitonic band, and ($i$,$j$) are the coordinate indices. Due to the selection rules, it is convenient to let $(i,j) = (+,-)$, corresponding to the circular polarizations $\sigma^+$ and $\sigma^-$.  

Figure~\ref{theory}(a)
shows the calculated velocity matrix elements $\langle \Psi_2 | (\bm{v}_e-\bm{v}_h) | \Psi_1 \rangle$ as a function of $P$ for the lowest two excitonic subbands (1$s$ and 2$p_-$) at 9\,T. We find that the velocity operator has nonzero elements in the $\sigma^+$ polarization for $P \sim 4\hbar/l_c$, while the conventional selection rule only allows the transition from 1$s$ to 2$p_-$ by absorbing $\sigma^-$ photons. This is because the Coulomb interaction at nonzero $P$ does not commute with the angular momentum operator, so excitonic states with different angular momenta are mixed. Figure\,\ref{theory}(b) shows calculated energy dispersions for the 1$s$ and 2$p_-$ subbands, the separation of which is the transition energy at a given $P$. From the dispersions, we determine the binding energy of the 1$s$ exciton at $P=0$ to be 13.9\,meV at 9\,T. At $P \approx$ 0, the transition energy agrees with the experimental  transition energy (6.36\,meV). At high temperatures, or large $P$, where the electron and hole are loosely bound, the transition energy is reduced, approaching the HH-CR energy (2.05\,meV) indicated by the blue arrow. Because the Coulomb-interaction-induced mixing of exciton states with different angular momenta is small, the only significant term of the conductivity is $\sigma_{--}$, which is related to the longitudinal conductivity, $\sigma_{xx}$, by $\sigma_{xx} = (1/2) \sigma_{--}$. Figures\,\ref{theory}(c) and \ref{theory}(d) show calculated real part of THz conductivity and permittivity, respectively, for magnetoexcitons at different temperatures for linear polarization at 9\,T, with an assumed pair density of 2.5\,$\times$\,10$^{10}$\,cm$^{-2}$.  These spectra qualitatively reproduce the experimentally observed temperature dependent spectra shown in Figs.\,\ref{various}(e) and \ref{various}(f). 

In summary, we presented results of a systematic experimental study of the exciton formation and ionization dynamics, including excitonic states with finite center-of-mass momenta, in photoexcited undoped GaAs QWs in magnetic fields up to 10\,T. 
In magnetic fields, we observed both the 1$s$--2$p_+$ and 1$s$--2$p_-$ transitions. We found that the 1$s$--2$p_-$ feature at 9\,T was extremely robust even under high excitation fluences, indicating magnetically enhanced stability of excitons.

We acknowledge support from the NSF (Grant Nos.~DMR-1310138 and OISE-0968405).  Work completed at Purdue was supported by the DOE, Office of BES, Division of Materials Sciences and Engineering under Award DE-SC0006671.  Work at TAMU was supported in part by AFOSR grant FA9550-15-1-0153. We thank R.\ Liu and M.\ Kira for useful discussions.


\end{document}